\documentclass[twocolumn,showpacs,preprintnumbers,amsmath,amssymb]{revtex4}

\usepackage{graphicx}
\usepackage{dcolumn}
\usepackage{bm}
\newcommand{\be}{\begin{equation}}
\newcommand{\ee}{\end{equation}}

\newcommand{\ber}{\begin{eqnarray}}
\newcommand{\eer}{\end{eqnarray}}

\newcommand{\bra}{\langle}
\newcommand{\ket}{\rangle}

\newcommand{\bs}[1]{\ensuremath{\boldsymbol{#1}}}
 
\begin{document}

\title{Superfluid to insulator phase transition in a unitary Fermi gas}
\author{Nir Barnea}
\email{nir@phys.huji.ac.il}
\affiliation{The Racah Institute of Physics, The Hebrew University,
91904 Jerusalem, Israel.\\
     Institute for Nuclear Theory, University of Washington, 98195 Seattle,
  Washington, USA}
\date{\today}

\begin{abstract}
We study the evolution of the energy gap in a unitary Fermi gas as a
function of temperature.
To this end we approximate the Fermi gas by the Hubbard lattice
Hamiltonian and solve using 
the dynamical mean-field approximation.
We have found that below the critical temperature, $T_c$, the system is a
superfluid and the energy gap is
decreasing monotonously. For temperatures above $T_c$ 
the system is an insulator and the corresponding energy gap is monotonously
increasing.  
\end{abstract}

\pacs{67.85.Lm, 05.30.Fk, 03.75.Ss}
\maketitle

{\it Introduction --}
Dilute Fermi gas, characterized with interparticle distance much larger than
the effective range but much smaller than the scattering length, 
$|a_s| \gg \sqrt[3]{\frac{3}{4\pi n}} \gg r_{eff} $, 
has been the
subject of intense theoretical and experimental research in the last few years
\cite{Giorgini07}. 
The interest in this system stems from its universal properties
that become independent of its
actual  
constituents as the scattering length diverges at unitarity
($|a_s|\longrightarrow\infty$) and depend only on the  
particle density. 
In the weak coupling regime ($a_s$ small and negative) the ground state of a
Fermi gas is a BCS superfluid. In the strong coupling limit ($a_s$ small
and positive) the fermions are bound in pairs that form a Bose-Einstein
condensate (BEC) for temperatures below the critical temperature. At unitarity
the system is in between 
these two 
limits 
and exhibits a distinct behavior which can be classified as a new type of 
superfluidity,  characterized by an admixture of bosonic and
fermionic features \cite{Bulgac06}.  

The phenomenon of superfluidity in Fermi systems is associated with the
occurrence of off-diagonal long range order, 
$\Delta_0=U\bra T c_{\uparrow}(0^+) c_{\downarrow}(0)\ket$, 
and the existence of
a gap $\Delta_{gap}$ in the single particle excitation spectrum.
In general, the order parameter $\Delta_0$ and the gap $\Delta_{gap}$ are
independent quantities. However, for weakly interacting fermions, in the BCS
regime, one finds $\Delta_{gap}=\Delta_0$. 

In an intriguing paper, Bulgac {\it et.} al. \cite{Bulgac08} used 
quantum Monte Carlo
technique to study the evolution of $\Delta_{gap}$ as a function of
temperature assuming a quasi-particle spectrum.
They have found that in contrast with the BCS theory, where $\Delta_{gap}$
vanishes at the critical temperature $T_c$, for unitary Fermi gas 
the magnitude of $\Delta_{gap}(T_c)$ is about two-thirds of the zero
temperature gap. Even more striking is the fact that at about $T_c$ the
gap's derivative flips sign, i.e. 
$\Delta_{gap}$ grows for $T>T_c$.

These results call for a better understanding of the excitation spectrum of the
finite temperature unitary Fermi gas. The aim of this work is to study these
aspects of the system using the
dynamic mean field approximation (DMFA) \cite{Georges92,Georges96}.
In the DMFA,
a lattice problem is mapped into a self-consistent embedded
impurity problem. In the limit of
infinite spatial dimensions $d\longrightarrow\infty$ this mapping becomes
exact due to the
localization of the self-energy \cite{Metzner89}. 
For 3D fermions which we consider here,
DMFA can be regarded as a simplification
in which a purely local self-energy is assumed,  
$\hat\Sigma(\bs{k},i\omega_n)\approx \hat\Sigma(i\omega_n)$ (hat denotes a
spinor matrix). 
The validity of this assumption for unitary Fermi gas has been examined in
\cite{Nir08a,Nir08b}, where the problem was approximated by the lattice Hubbard
Hamiltonian, 
\begin{equation}\label{H_lattice}
  H = -t \sum_{\sigma \bs{n}\bs{n}'} D_{\bs{n n}'} 
                       \psi^{\dagger}_{\bs{n} \sigma}
                       \psi_{\bs{n}' \sigma}
    +  U \sum_{\bs{n}} 
                   \psi^{\dagger}_{\bs{n}\, \uparrow}
                   \psi^{       }_{\bs{n}\, \uparrow}
                   \psi^{\dagger}_{\bs{n}\, \downarrow}
                   \psi^{       }_{\bs{n}\, \downarrow}\;,
\end{equation}
and the continuum limit was realized by reducing the lattice filling to zero.
It was found \cite{Nir08b} that the DMFA results agree remarkably well with
those of full quantum Monte-Carlo simulations (QMC) 
\cite{GFMC04,Giorgini04,Carlson03,Carlson05},
yielding  
$\xi\approx 0.44$ for the ratio between the energy per particle of the
interacting 
and free systems ($E/N=\xi E_{FG}$), and $\Delta_0 \approx 0.64 E_F$.
In this letter we use the DMFA to study the finite temperature
excitation spectrum of the unitary Fermi gas.

{\it DMFA -}
Using the Nambu formalism, the DMFA single-site impurity effective action takes
the form 
\begin{eqnarray}\label{S_eff}
 S_{eff} & = & -\int_0^{\beta}d\tau \int_0^{\beta}d\tau'
 \Psi^{\dagger}(\tau)\hat{\cal G}^{-1}_0(\tau-\tau')  
 \Psi^{}(\tau') 
\cr & & 
 - U \int_0^{\beta}d\tau \, 
  c^{\dagger}_{\uparrow  }(\tau)c^{}_{\uparrow  }(\tau)
  c^{\dagger}_{\downarrow}(\tau)c^{}_{\downarrow}(\tau) \;,
\end{eqnarray}
where $\beta=1/T$ is the inverse temperature, $\Psi^{\dagger} \equiv
(c_{\uparrow}^{\dagger},c_{\downarrow})$ 
are the Nambu spinors, and the bath's
Green's function $\hat{\cal G}_0$  
is determined 
through the self-consistency condition that the impurity Green's function 
$ \hat{\cal G}(\tau) \equiv 
      -\bra T \Psi^{}_{i}(\tau) \Psi^{\dagger}_{i}(0) \ket_{S_{eff}} $
coincides with the site-diagonal lattice Green's function calculated
with the self-energy 
$ \hat \Sigma(i\omega_n)=\hat {\cal G}^{-1}_0(i\omega_n)
                       -\hat {\cal G}^{-1}(i\omega_n)\,. $

We use the direct diagonalization method of
Caffarel and Krauth \cite{Caffarel94} to solve the DMFA. In this approach the
impurity action 
is mapped into the Anderson Hamiltonian
\begin{eqnarray}\label{H_And}
\lefteqn{{\cal H}_{And} 
    =  \sum_{l,\sigma} \tilde\epsilon_l a^{\dagger}_{l
  \sigma}a^{}_{l \sigma}    
+ \sum_{l,\sigma} \tilde V_l ( a^{\dagger}_{l \sigma} c_{\sigma}
  +c^{\dagger}_{\sigma} a^{}_{l \sigma} ) }
\cr & + &
 \sum_{l,\sigma} \tilde D_l ( a^{\dagger}_{l \sigma} c^{\dagger}_{-\sigma}
  +c^{}_{-\sigma} a^{}_{l \sigma} )
-\mu \sum_{\sigma} c^{\dagger}_{\sigma}c^{}_{\sigma}
+ U c^{\dagger}_{\uparrow} c_{\uparrow} 
    c^{\dagger}_{\downarrow} c_{\downarrow} \;, 
\cr & &
\end{eqnarray}
where the interaction of the fermionic field
$c_{\sigma}$ with the auxiliary bath fermions $a_{l\sigma}$
generate $\hat{\cal G}_0$. 
This goal is achieved by choosing the parameters of the Anderson model 
$\tilde \epsilon_l, \tilde V_l, \tilde D_l$ 
to minimize the difference between the $\hat {\cal G}_0$ and 
$\hat {\cal G}_0^{And}$.
In this work we use $4-5$ auxiliary fermionic fields. For lattice
filling $n=0.1$ which we consider here, taking this number of auxiliary fields
yields an accuracy 
of about $1\%$ for the thermodynamic quantities \cite{Nir08b}.

{\it The Excitation Spectrum -}
The determination of real-frequency quantities such as the spectral function
or the excitation spectrum faces severe limitations in QMC simulations where
only imaginary time/frequency data are obtained directly. Trying to overcome
this limitation Bulgac {\it et.} al. \cite{Bulgac08} have calculated, 
using a QMC simulation, the susceptibility
\begin{equation}\label{chi_def}
  \chi(\bs{k}) = - \int_0^{\beta} d\tau G(\bs{k},\tau) 
               = - \frac{2}{\beta} \sum \frac{1}{i \omega_n}
                  G(\bs{k},i \omega_n ) \;,
\end{equation}
where $G(\bs{k},\tau)$ is the Green's
function and $\omega_n={(2n+1)\pi}/{\beta}$ are the Matsubara frequencies. 
For an independent-(quasi) particle spectrum the response (\ref{chi_def})
can be easily evaluated,
\begin{equation}\label{chi_qp}
   \chi(\bs{k}) = \frac{1}{E_{\bs{k}}}
                  \frac{e^{\beta E_{\bs{k}}}-1} {e^{\beta E_{\bs{k}}}+1} \;,
\end{equation}
where $E_{\bs{k}}$ are the single-(quasi)particle excitation energies.
Exploiting this observation, they have fitted the calculated susceptibility
to the formula (\ref{chi_qp}) assuming, given the chemical potential
$\mu$, the spectrum
\begin{equation}\label{E_qp}
E_{\bs{k}}^{qp}=\sqrt{(\alpha_{qp} \epsilon_{\bs{k}} + \Sigma_{qp} -\mu)^2
                   +\Delta_{qp}^2} \;,
\end{equation}
treating 
$\alpha_{qp}, \Sigma_{qp}, \Delta_{qp}$ as free parameters. 
These parameters stand for the effective mass $m^*=m/\alpha$, mean field
potential $\Sigma_{qp}$, and the ``pairing'' gap $\Delta_{gap}=\Delta_{qp}$. 
$\epsilon_{\bs{k}}$ is the free-particle kinetic energy.

Solving the DMFA equations with the direct diagonalization method 
\cite{Caffarel94} the spectral function can be obtained
directly from the impurity model but in the form of a set of delta functions.
Since we are limited to a finite and rather small number of orbitals in the
effective bath it is difficult to extract quantitative information from it. 
Consequently one has to adopt a different strategy in order to
study the real frequency properties of the system. 

Consider the occupation probability  
\begin{equation}\label{f_def}
  f(\bs{k}) = G(\bs{k},0^+) = \frac{1}{\beta} \sum e^{i\omega_n 0^{+}}
                  G(\bs{k},i \omega_n ) \;,
\end{equation}
the Green's function derivative at $\tau=0^+$
\begin{equation}\label{zeta_def}
  \zeta(\bs{k}) = \left. \frac{d G(\bs{k},\tau)}{d \tau}\right|_{\tau=0^+}
                = \frac{1}{\beta} \sum e^{i\omega_n 0^{+}}i\omega_n
                  G(\bs{k},i \omega_n ) \;,
\end{equation}
and the susceptibility $\chi(\bs{k})$ defined above (\ref{chi_def}). For  
an independent-(quasi)particle Green's function, 
\begin{equation}
  G_{qp}(\bs{k},i\omega_n) = 
      \frac{i\omega_n -\mu+\epsilon_{\bs{k}}+\Sigma_{qp}}
           {(i\omega_n-E_{\bs{k}})(i\omega_n+E_{\bs{k}})}\;,
\end{equation}
these quantities
can be manipulated to yield the relation
\begin{equation}\label{e_qp_dmfa}
    E_{\bs{k}} = \sqrt{-\frac{1}{\chi(\bs{k})}
          \left[2\zeta(\bs{k})+\frac{2 f(\bs{k}) -1}{\chi(\bs{k})}\right]}\;.
\end{equation}
This procedure is a generalization of \cite{Bulgac08} that avoids, however,
the need to invert Eq. (\ref{chi_qp}). 
In the DMFA $\chi(\bs{k}), f(\bs{k}), \zeta(\bs{k})$ can be easily calculated 
through the Matsubara sums (\ref{chi_def}), (\ref{f_def}), and
(\ref{zeta_def}). Once we have performed these sums the value of $E_{\bs{k}}$
can be evaluated (\ref{e_qp_dmfa}), regardless of the original assumption about
the 
nature of the excitation spectrum. Strictly speaking, only for a limited number
of cases the identification of Eq. (\ref{e_qp_dmfa}) with
the quasi-particle excitation spectra is exact. 
Nevertheless, in the following we shall refer to it as the
quasi-particle excitation spectrum.

\begin{figure}[ht]
\includegraphics[height=6cm]{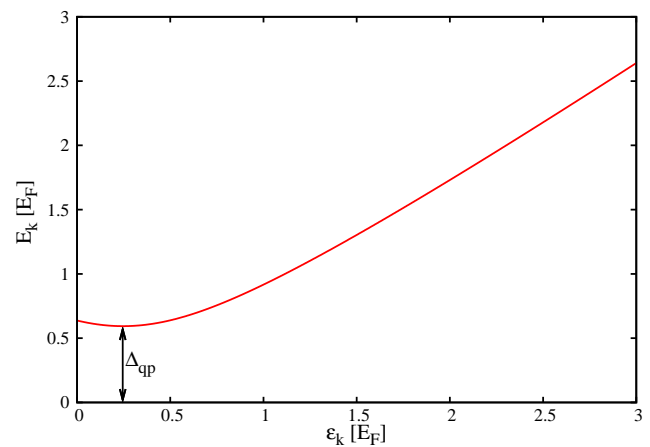}%
\caption{\label{fig:e_qp} (Color online)
The quasi-particle spectrum as calculated from Eq. (\ref{e_qp_dmfa}) for a
unitary 
Fermi gas at $T=0.38 E_F$ ($T_c \approx 0.16 E_F$) and lattice filling
$n=0.1$. The minimum in the graph 
corresponds to the quasi-particle gap $\Delta_{qp}$.
}
\end{figure}
\begin{figure}[ht]
\includegraphics[height=6cm]{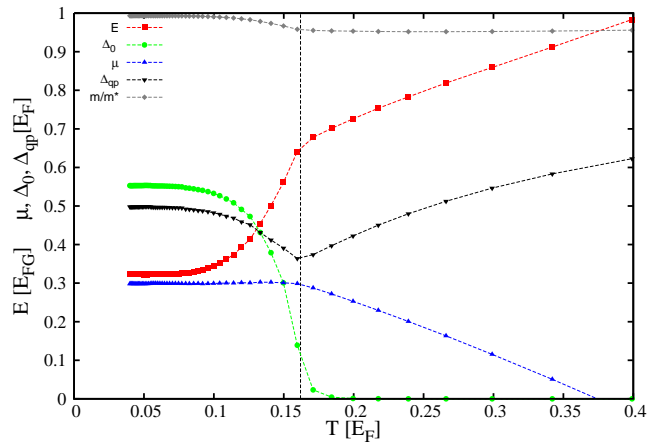}%
\caption{\label{fig:tdf} (Color online)
The phase transition from a superfluid to a normal
phase in a unitary Fermi gas. 
The energy per particle $E$ is shown by squares, the order parameter
$\Delta_0$ by circles, the chemical potential $\mu$ by up triangles, the
quasi-particle gap $\Delta_{qp}$ by down triangles, and the bare to effective
mass ratio $m/m^*$ by diamonds.
}
\end{figure}

In Fig. \ref{fig:e_qp} the quasi-particle
energy (\ref{e_qp_dmfa}) is plotted as a function of the free-particle kinetic
energy $\epsilon_{\bs{k}}$ for lattice filling $n=0.1$, at temperature $T=0.38
E_F$, beyond the phase transition temperature
which in our calculation is $T_c\approx 0.16 E_F$. 
From the figure it can be seen
that $E_{\bs{k}}$ exhibits an excitation spectrum typical for a gapped system,
to which we shall refer as an insulator.
The gap can be evaluated directly from the graph as the minimum of
$E_{\bs{k}}$. Fitting the empirical formula (\ref{E_qp}) one can reproduce the
excitation spectrum (\ref{e_qp_dmfa}) very accurately. Using this procedure we
get an estimate for $\Delta_{qp}$ even when the minimum of (\ref{e_qp_dmfa}) is
outside the band and we also get an estimate for the effective mass.
In Fig. \ref{fig:tdf} we plot for a unitary Fermi gas the gap $\Delta_{qp}$ and
$m/m^*$ as a function of $T$ at lattice filling $n=0.1$. Also plotted are
the energy per particle $E$, the chemical potential $\mu$, and the order
parameter $\Delta_0$. The phase transition from a superfluid to 
a normal phase is associated with the vanishing of the order parameter
and a jump in the heat capacity. Such transition is easily located in Fig.  
\ref{fig:tdf} at $T\approx 0.16 E_F$. Following the evolution of $\Delta_{qp}$
with $T$, we see that at low temperatures $\Delta_{qp}$ is a decreasing
function of temperatures up to $T_c$. At which an abrupt change is observed
and for $T\geq T_c$ the gap is increasing with $T$.
This discontinuity in the derivative $d\Delta_{qp}/dT$ is a clear indication
that the quasi-particle gap has a different meaning in the two phases. For
$T\leq T_c$ it can be associated with the superfluid gap,
$\Delta_{gap}$. This interpretation, however, is lost for $T\geq T_c$.

\begin{figure}[ht]
\includegraphics[height=6cm]{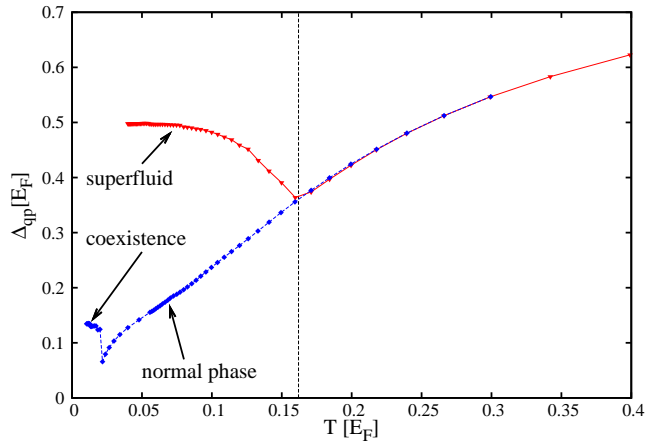}%
\caption{\label{fig:delta_qp_bcsfl} (Color online)
The quasi-particle gap as a function of $T$. The thermodynamic stable 
solution is shown by triangles. The frustrated superfluid
solution (the normal solution) is presented by diamonds. }
\end{figure}

A better understanding of the quasi-particle gap at $T > T_c$ can be
achieved by frustrating the superfluid phase at $T\leq T_c$. Within the
DMFA this goal can be easily achieved by forcing particle number conservation
in the effective impurity action, i.e. by setting $\tilde D_l=0$ in
(\ref{H_And}).  
In Fig. \ref{fig:delta_qp_bcsfl} we present
the results of such calculation for the quasi-particle gap. 
At very low temperatures $T\leq T_{pairing}\approx 0.02 E_F$ the frustrated
solution exhibits 
a gap of about $0.13 E_F$, which corresponds to the coexistence region
of a metal phase and a pairing (insulator) phase. This phase transition
in the normal, unstable, phase around the unitarity limit was 
identified at $T=0$ by Keller {\it et.} al. \cite{Keller01} for an infinite
dimensional system. 
Toschi {\it et.} al. \cite{Toschi05b} have studied the finite temperature
phase diagram of the frustrated solution, and found that although there is a
smooth transition between the metal and
insulator phases at temperature above the metal-pairing 
critical point (which is much lower than $T_c$) the properties of the system 
depend strongly on the strength of the coupling constant.
Above $T_{pairing}$ we see a drop in
the normal phase gap which then grows almost linearly with $T$ until $T_c$.
It is interesting to note that the decreasing superfluid gap and the 
increasing normal phase gap coincide at $T_c$. 
It is evident that
for $T > T_c$ the thermodynamic stable gap $\Delta_{qp}$ shifts from
describing the superfluid solution into the 
normal phase solution. 
From the figure it seems that the gap saturates at higher temperatures. Being
limited by the band width there is no point in carrying our calculations to
higher values of $T$.

\begin{figure}
\includegraphics[height=6cm]{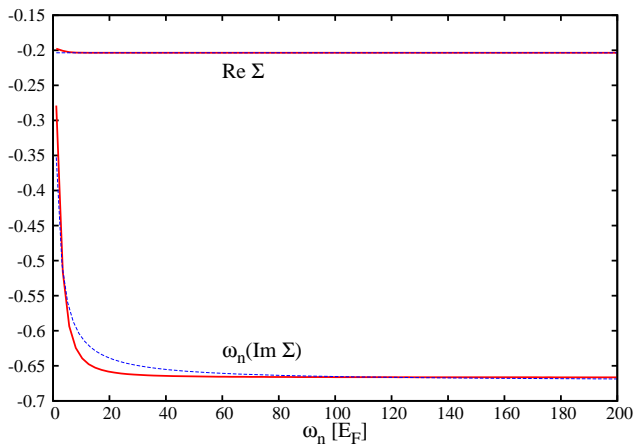}%
\caption{\label{fig:selfe} (Color online)
The self-energy of a unitary Fermi gas at 
$T=0.365 E_F $ and lattice filling $n=0.1$. The real part of $\Sigma$ and
imaginary part multiplied by $\omega_n$ are plotted with thick lines. The
dashed line is the simplified function (\ref{se_fit}) with best fit parameters
$\Sigma_0=-0.20  E_F$, $\eta_0=0.82 E_F$, and $\theta_0=1.04 E_F$.
}
\end{figure}

We can achieve further insight into the behavior of the unitary Fermi gas by
inspecting the self-energy, $\Sigma(i\omega_n)$, at $T \geq T_c$, see
Fig. \ref{fig:selfe}.  
It can be seen that the real part of $\Sigma(i\omega_n)$ is essentially
constant. The imaginary part decrease asymptotically as $1/\omega_n$ but tends
towards 
a finite value as $\omega_n \longrightarrow 0$. 
Consequently $\Sigma(\omega_n)$ can be roughly represented
in the form,
\be \label{se_fit}
   \Sigma(i\omega_n) \approx \Sigma_{0}+ \frac{\eta_0^2}{i\omega_n+i\theta_0}
\ee
where $\eta_0$ characterize the asymptotic behavior of $\rm{Im} \Sigma$ and 
$\theta_0$ is used to model the low frequency behavior.
The dashed line in Fig. \ref{fig:selfe} is a fit of (\ref{se_fit}) to 
the calculated self-energy at $T=0.365 E_F$. The best fit parameters are
$\Sigma_0=-0.20  E_F$, $\eta_0=0.82 E_F$, and $\theta_0=1.04 E_F$.
Using (\ref{se_fit}) we get an analytic model for the thermal Green's
function. 
The upper plane part of this Green's function can be related to the
retarded Green's function yielding,
\be
    G^R(\omega,\bs{k})=\frac{1}{\omega+\mu-\epsilon_{k}-\Sigma_0
                                -\frac{\eta_0^2}{\omega+i\theta_0}} \;.
\ee
In this model the Green's function contains two poles
\begin{equation}
    \omega_{\pm}=\frac{
    e_{\bs{k}}-i\theta_0
     \pm \sqrt{(e_{\bs{k}}+i\theta_0)^2+4\eta_0^2}
     }{2}
\end{equation}
where $e_{\bs{k}}=\epsilon_{\bs{k}}+\Sigma_0-\mu$,
and can be written as
\be
    G^R(\omega,\bs{k})=\frac{\omega+i\theta_0}{\omega_{+}-\omega_{-}}
    \left(\frac{1}{\omega-\omega_{+}}-\frac{1}{\omega-\omega_{-}}
    \right) \;.
\ee
The two poles come close to each other as $\epsilon_{\bs{k}}$ approach 
the Fermi surface. At $\epsilon_{\bs{k}}=\mu-\Sigma_0$ we get a simple
expression for the gap,
$ {\Delta\omega = \omega_{+}-\omega_{-}=\sqrt{4\eta_0^2-\theta_0^2}
        }$. 
This expression implies that for $\eta_0 > \theta_0/2$ there is a
real gap in the spectrum and the system can be characterized as an
insulator whereas for $\eta_0 \leq \theta_0/2$ there is no gap and we
may characterize the system as a Fermi liquid.
For the example in Fig. \ref{fig:selfe}, 
$\Delta_{gap}=\Delta\omega/2 \approx 0.63 E_F$ 
in nice agreement with the quasi-particle gap 
$\Delta_{qp}\approx 0.6 E_F$, see Fig. \ref{fig:delta_qp_bcsfl}.
Inspecting the residues $R_{\pm}$ it is clearly seen that as
$\epsilon_{\bs{k}}$ pass trough the Fermi surface the power is shifted from
$\omega_-$ to $\omega_+$,
\begin{eqnarray}
  \epsilon_{\bs{k}}\longrightarrow -\infty 
    & R_{+}\rightarrow 0 &  R_{-}\rightarrow 1
\cr
  \epsilon_{\bs{k}}\longrightarrow +\infty 
    &  R_{+}\rightarrow 1  & R_{-}\rightarrow 0 \;.
\end{eqnarray}
This simple, but rather exact, parameterization implies the existence of a gap
in the excitation spectrum and 
supports the conclusions we drew from the quasi-particle spectrum
(\ref{e_qp_dmfa}). 

{\it Conclusions -}
Using the dynamic mean field approximation we have studied  for a unitary
Fermi gas the evolution of the 
quasi-particle gap with 
temperature. We have found, in accordance with QMC
calculations \cite{Bulgac08}, that in the superfluid phase the gap decreases up
to $T_c$ and 
then starts to rise. We have shown that at $T_c$ there is a sharp transition in
the gap's slope. This transition is associated with a shift from the
superfluid to the normal phase gap. We have demonstrated, by frustrating the
superfluid solution, that the normal phase
insulator gap is much smaller than the superfluid gap at low temperatures. 
The insulator gap grows with increasing temperature and
the two gaps coincide just at $T_c$. The connection between 
the lost of coherence and the gap crossing is not yet clear. 


I wish to thank G. F. Bertsch, A. Bulgac, D. Gazit, and P. Magierski
for 
useful discussions and help 
during the preparation of this work.
This work was supported by the Department of Energy Grant
No. DE-FG02-00ER41132. 

\end{document}